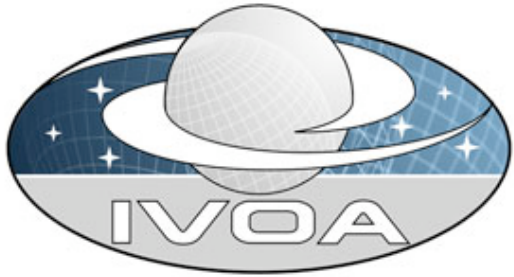

# IVOA Support Interfaces
# Version 1.00-20110531

## IVOA WG Recommendation 2011 May 31

**Editors:**
> Matthew Graham
> Guy Rixon

**Author(s):**
> Grid and Web Services Working Group

---

## Abstract

This document describes the minimum interface that a (SOAP- or REST-based) web service requires to participate in the IVOA. Note that this is not required of standard VO services developed prior to this specification, although uptake is strongly encouraged on any subsequent revision. All new standard VO services, however, must feature a VOSI-compliant interface.

## Status of this Document

Recommendation

This document has been produced by the IVOA Grid and Web Services Working Group. It has been reviewed by IVOA Members and other interested parties, and has been endorsed by the IVOA Executive Committee as an IVOA Recommendation. It is a stable document and may be used as reference material or cited as a normative reference from another document. IVOA's role in making the Recommendation is to draw attention to the specification and to promote its widespread deployment. This enhances the functionality and interoperability inside the Astronomical Community.

A list of current IVOA Recommendations and other technical documents can be found at http://www.ivoa.net/Documents/.

# Acknowledgments

This document has been developed with support from the National Science Foundation's Information Technology Research Program under Cooperative Agreement AST0122449 with The Johns Hopkins University, from the UK Particle Physics and Astronomy Research Council (PPARC), from the European Commission's (EC) Sixth Framework Programme via the Optical Infrared Coordination Network (OPTICON), and from EC's Seventh Framework Programme via its eInfrastructure Science Repositories initiative.

This work is based on discussions and actions from the 2003 IVOA meeting in Strasbourg and further discussions on registry functionality at JHU late in 2003. Later inputs came from a local meeting at JHU in Sept. 2004. William O'Mullane and Ani Thakar were the editors and primary authors for these early versions.

The decision to split the interfaces into a mandatory set and optional logging interfaces was taken by GWS-WG at the IVOA meeting of May 2006.

# Conformance related definitions

The words "MUST", "SHALL", "SHOULD", "MAY", "RECOMMENDED", and "OPTIONAL" (in upper or lower case) used in this document are to be interpreted as described in IETF standard, RFC 2119 [0].

The **Virtual Observatory (VO)** is a general term for a collection of federated resources that can be used to conduct astronomical research, education, and outreach. The **International Virtual Observatory Alliance (IVOA)** is a global collaboration of separately funded projects to develop standards and infrastructure that enable VO applications. The International Virtual Observatory (IVO) application is an application that takes advantage of IVOA standards and infrastructure to provide some VO service.

# Contents

# 1. Introduction

The web services that comprise much of the Virtual Observatory (VO) come in two forms: SOAP-based (Simple Object Access Protocol, [1]) ones such as footprint and spectrum services [2], SkyNodes and Open SkyQuery [3], registry interfaces [4] and CDS access [5]; and RESTful (REpresentational State Transfer, [6]) ones such as TAP [7] and other second generation data access (DAL) services, and VOSpace 2.0 [8]. This document describes a set of common basic functions that all these services should provide in the form of a standard support interface in order to support the effective management of the VO. It is agreed that VO service standards previous to VOSI may not be forced to retrospectively implement VOSI (although that should be encouraged). Nonetheless, all new VO service standards (or updated existing ones) must enforce the VOSI implementation.

The IVOA Web Services Basic Profile [9] mandates that a compliant SOAP-based web service should have the interface defined in this specification, as expressed using the standard WSDL (Web Services Description Language, [10]) format. For RESTful services, the requirement for the support interface is stated in the specification for each kind of service. A contract for a RESTful service may specify extra constraints (e.g., on the form of the URIs) for the support interface. Such a contract might be expressed using the WADL (Web Application Description Language, [15]) format.

## 1.1 The role in the IVOA Architecture

The IVOA Architecture [11] provides a high-level view of how IVOA standards work together to connect users and applications with providers of data and services, as depicted in the diagram in Fig. 1.

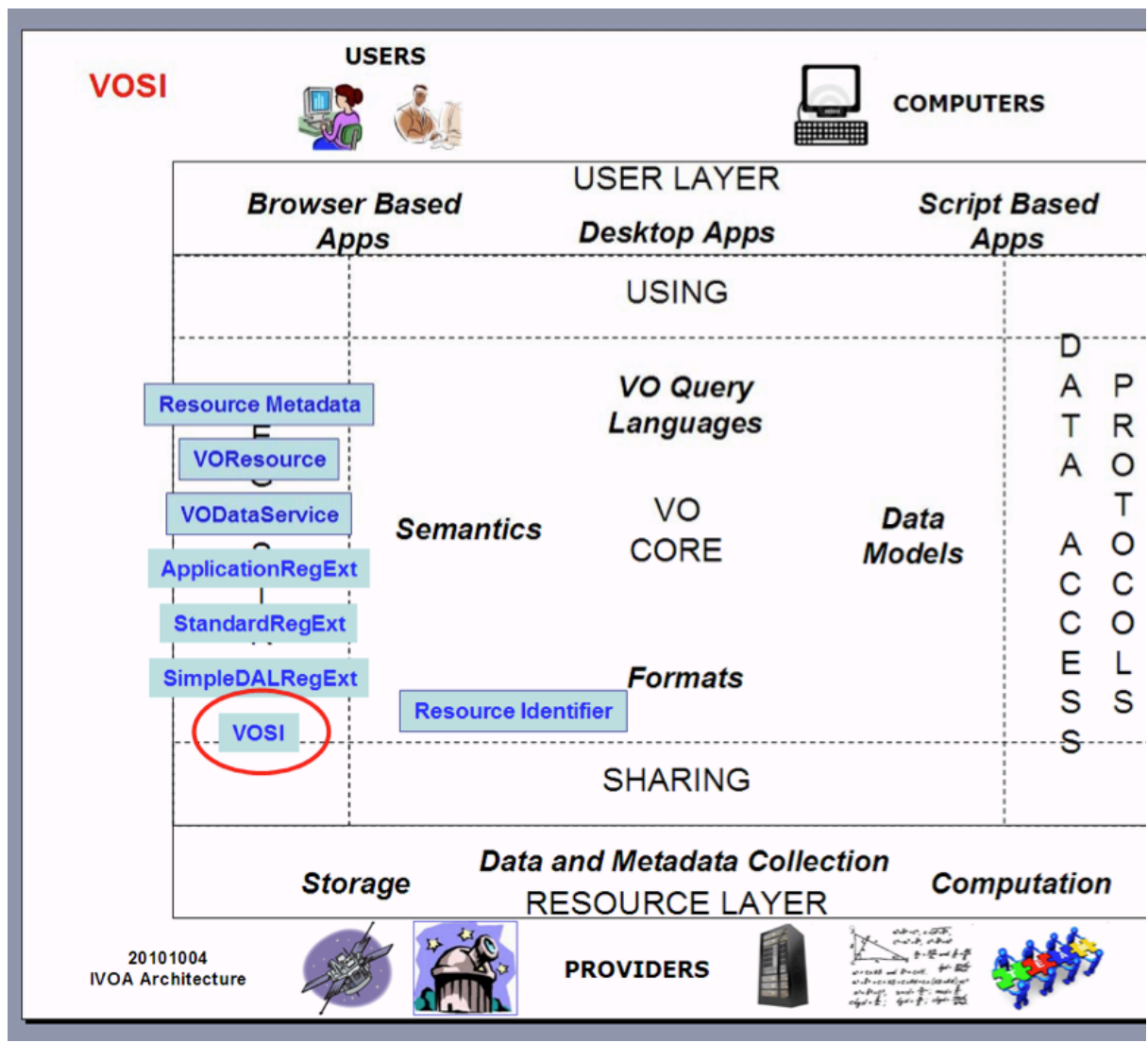

**Figure 1. VOSI in the IVOA Architecture.** VOSI is the standard that defines the basic functions that all VO services should provide in order to support management of the VO.

In this architecture, users employ a variety of tools (from the User Layer) to discover and access archives and services--that is, resources--of interest (represented in the Resource Layer). A registry plays a role in discovery by harvesting metadata that describe archives and services and making them searchable from a central service. The VOSI interface provides a means for a service to provide some of this metadata itself; this allows a registry to pull the metadata from the service rather than relying on a human to provide it (e.g. by typing the data into a registration form manually). This mechanism can make it easier to collect highly detailed metadata (e.g. descriptions of columns in a catalog) that might not be practically provided otherwise. As some of this metadata describes the service interface and how it behaves, other applications can use this information for controlling how they use the service. Even when the service is "discovered" through some means other than a registry, an application can still understand how to use the service by querying for this information directly. (See Appendix B for a more detailed description of this use case.)

Once a user discovers data and services of interest, she will want to engage them in an analysis process. Success requires that the selected services are actually up and running properly as a down service can cause automated processing to fail completely. Registry and workflow services can assist with this by tracking the availability of services and alerting users about downtime. We envision that VOSI will allow VO projects to better track the overall health of the VO ecosystem.

## 2. Interface bindings

The standard interface returns metadata without changing the state of the service with which it is associated. This could, in principle, be implemented in any of three ways:

1. As extra SOAP operations on an existing SOAP endpoint of the service with which it is associated. This would be a 'SOAP binding' of VOSI.
2. As SOAP operations on a separate SOAP endpoint. This would be an alternate form of the SOAP binding.
3. As web resources with distinct URLs, without using the SOAP protocol. This is the 'REST binding' for the standard interface.

This standard requires the REST binding of VOSI even when applied to services that otherwise use SOAP. No details of the SOAP binding are given in this version of the standard.

In the REST binding, the support interfaces shall have distinct URLs in the HTTP scheme and shall be accessible by the GET operation in the HTTP protocol. The response to an HTTP POST, PUT or DELETE to these resources is not defined by this specification. However, if an implementation has no special action to perform for these requests, the normal response would be a HTTP 405 "Method not allowed" status code.

The endpoints and interface types for the support interface shall be defined in the service's registration using one *Capability* element for each interface. The values of the *standardID* attribute for these *Capability*s are given in section 4.

When using the REST binding, any HTTP URLs may be used. The client must find the appropriate URLs from the service's entry in the VO registry and, in general, should not try and infer the URLs from any other URLs for that service. However, standards for specific services may put extra constraints on the form of the URLs.

## 3. Metadata specification

There are various classes of metadata that might be returned by a service through its standard interface:

- those describing its functional capabilities
- those describing its operational behaviour - availability, reliability, etc.
- those describing tabular data handled by the service
- those describing other aspects of the service

This section defines how each of these classes is represented. The following typographic convention is used to represent a XML element defined within a particular namespace:

```
{http://some.name.space}elementName
```

For example, *{http://www.ivoa.net/xml/VOResource/v1.0}resource* indicates a XML element named *resource* that is described in the XML schema associated with the 'http://www.ivoa.net/xml/VOResource/v1.0' namespace - in this case, this would be VOResource.xsd [12].

## 3.1 Capability metadata

> **Note:**
>> 'Capability' is unfortunately an overloaded term in the VO referring to both a functional aspect of a service and also particular pieces of metadata defined by various XML schema. When referring to an XML element called 'capability', it shall be be put in italics. Its parent namespace may also be included (using the syntax described above) if it is ambiguous which XML schema is being referred to.

This interface provides the service metadata in the form of a list of *Capability* descriptions. Each of these descriptions is an XML element that:

- states that the service provides a particular, IVOA-standard function;
- lists the interfaces for invoking that function;
- records any details of the implementation of the function that are not defined as default or constant in the standard for that function.

For example, one *Capability* might describe a cone search function and another the TAP implementation but these two might well apply to the same service.

An entry for a service in the resource registry - i.e., its VOResource - contains the Dublin Core resource metadata (identifier, curation information, content description, etc.) followed by the service's capability descriptions (expressed as a series of *{http://www.ivoa.net/xml/VOResource/v1.0}Capability* elements). Effectively, the resource metadata describes the service to human users and the capability list describes it to software. Therefore, the latter list has two uses:

- it may be read by a client application to find out how to invoke the service. This presumes that the service has been already been selected and the VOSI endpoint located.
- it may be read by the registry itself to compile the registry entry for the service. In this case, the resource metadata are entered into the registry directly and the service metadata are then read from the service. Since the service implementation usually knows its capabilities, this removes the need for a human to type them into the registry.

The service metadata shall be represented as an XML document with the root element *{http://www.ivoa.net/xml/VOSICapabilities/v1.0}capabilities*. (See [Appendix A.1](#) for the definition of the VOSICapabilities XML schema.) This element must contain one or more child *capability* elements that describe the capabilities of the service. Given that the *capability* element is defined to be of type *{http://www.ivoa.net/xml/VOResource/v1.0}Capability*, a *capability* element may be represented by a legal sub-type of *{http://www.ivoa.net/xml/VOResource/v1.0}Capability*, in which case, the *capability* element must use an *xsi:type* attribute to identify the sub-type (see section 2.2.1 of

[12]]).

> **Note:**
>
> The value of the *capability* element's standardID attribute is used to indicate the service's support for particular standard protocols (such as Simple Image Access, Simple Cone Search, etc.). In the case of some protocols, the support for the standard is further characterized by additional metadata provided by a standard XML schema extension of *Capability* for that protocol. The extension metadata is enabled by adding a *xsi:type* attribute to the *capability* element set to the *Capability* sub-type value defined in the extension schema for that protocol (see example below).

The VOResource list of capabilities should include capabilities describing VOSI endpoints as specified in section 4.

In the REST binding, the service metadata shall be a single web resource with a registered URL. The date and time at which the metadata last changed shall be obtained from the *Last-Modified* HTTP Header keyword sent in the response to a GET or HEAD request to the registered URI.

All VO services should provide this interface.

## 3.2 Non-service metadata (non-normative)

There may be other metadata associated with a service than the capability metadata described above.

- Every service has the Dublin Core resource metadata [13].
- Some services are associated with registered applications.
- Some services are associated with registered data collections.

None of these are explicitly provided for in this version of VOSI. Some might be covered in later versions of VOSI.

## 3.3 Availability metadata

This interface indicates whether the service is operable and the reliability of the service for extended and scheduled requests. The availability shall be represented as an XML document in which the root element is *{http://www.ivoa.net/xml/Availability /v1.0}availability*. This element shall contain child elements providing the following information.

- *available* - whether the service is currently accepting requests
- *upSince* - duration for which the service has been continuously available
- *downAt* - the instant at which the service is next scheduled to be unavailable
- *backAt* - the instant at which the service is scheduled to become available again after down time;
- *note* - textual note, e.g. explaining the reason for unava\ilaility.

The elements *upSince*, *downAt*, *backAt* and *note* are optional. The *available* element is mandatory. There may be more than one *note* element.

The XML document shall conform to the schema given in appendix A.2 of this specification.

When reporting availability, the service should do a good check on its underlying parts to see if it is still operational and not just make a simple return from a web server, e.g., if it relies on a database it should check that the database is still up. If any of these checks fail, the service should set *available* to false in the availability output.

If a service is to be online but unavailable for work (e.g., when a service with a work queue intends to shut down after draining the queue) then the service should set *available* to false.

There are no special elements in the availability document for the contact details of the service operator. These details may be given as a *note* element if they are known to the service.

In the REST binding, the availability shall be a single web resource with a registered URL

All VO services shall provide this interface.

## 3.4 Table metadata

Some services deal with tabular data. These data may be the target of ADQL queries, as in TAP [7], or they may be the output of other operations, as in SIAP queries. In each case, it is useful if the service describes the details of the tables concerned. It is more useful if this description can be captured in the resource registry.

The *VODataService* standard [14] defines XML elements for describing a set of tables. These elements can be included in a resource document for a service.

A service which uses tables in its interface should define a VOSI endpoint from which table metadata can be read. The table metadata shall be represented as an XML document of which the root element is of type *{http://www.ivoa.net/xml/VODataService /v1.1}TableSet*. This element may contain any mix of elements allowed by the VODataService XML schema.

In the REST binding, the table metadata shall be a single web resource with a registered URL.

# 4. Registration of VOSI endpoints

The endpoints for the service and availability metadata shall be included in the registration of each service that provides them.

| Endpoint type | standardID value |
| --- | --- |
| availability | ivo://ivoa.net/std/VOSI#availability |
| capabilities | ivo://ivoa.net/std/VOSI#capabilities |
| tables | ivo://ivoa.net/std/VOSI#tables |

An availability endpoint shall be represented by an element named *capability*, of type

*{http://www.ivoa.net/xml/VOResource/v1.0}Capability* (defined by the standard VOResource XML schema [12]). The value of the *standardID* attribute of the *capability* shall be *ivo://ivoa.net/std/VOSI#availability*.

A capabilities endpoint should be represented by an element named *capability*, of type *{http://www.ivoa.net/xml/VOResource/v1.0}Capability*. If such a *capability* is provided then the value of the *standardID* attribute must be *ivo://ivoa.net/std/VOSI#capabilities*.

A tables endpoint should be represented by an element named *capability*, of type *{http://www.ivoa.net/xml/VOResource/v1.0}Capability*. If such a *capability* is provided then the value of the *standardID* attribute must be *ivo://ivoa.net/std/VOSI#tables*.

With all three VOSI functions, the *Capability* element that describes the function must contain an interface element of a type semantically appropriate for the binding of the function to the service; the *accessURL* element within the interface element indicates the endpoint for the VOSI function. For the REST binding, this *accessURL* element must set the *use* attribute to "full". Furthermore, for the REST binding, this document recommends using the *{http://www.ivoa.net/xml/VODataService/v1.1}ParamHTTP* interface type to encode VOSI endpoints (see example given in section 2.1).

# 5. Example VOSI responses

**Example 1:**

A sample response from a capabilities resource describing an SIA service.

```
<?xml version="1.0" encoding="UTF-8"?>

<vosi:capabilities xmlns:vosi="http://www.ivoa.net/xml/VOSICapabilities/v1.0"

                   xmlns:vr="http://www.ivoa.net/xml/VOResource/v1.0"

                   xmlns:vs="http://www.ivoa.net/xml/VODataService/v1.0"

                   xmlns:sia="http://www.ivoa.net/xml/SIA/v1.0"

                   xmlns:xsi="http://www.w3.org/2001/XMLSchema-instance"

                   xsi:schemaLocation="http://www.ivoa.net/xml/VOSI/v1.0

                                       http://www.ivoa.net/xml/VOSI/v1.0

                                       http://www.ivoa.net/xml/VOResource/v1.

                                       http://www.ivoa.net/xml/VOResource/v1.

                                       http://www.ivoa.net/xml/VODataService/

                                       http://www.ivoa.net/xml/VODataService/

                                       http://www.ivoa.net/xml/SIA/v1.0

                                       http://www.ivoa.net/xml/SIA/v1.0">

    <!-- a generic capability (for custom, non-standard interfaces) -->

    <capability>
```

```xml
    <interface xsi:type="vr:WebBrowser">

      <accessURL use="full"> http://adil.ncsa.uiuc.edu/siaform.html </accessURL>

    </interface>

  </capability>

  <!-- the SIA capability -->

  <capability xsi:type="sia:SimpleImageAccess" standardID="ivo://ivoa.net/std/SI

    <interface xsi:type="vs:ParamHTTP" role="std">

      <accessURL> http://adil.ncsa.uiuc.edu/cgi-bin/voimquery?survey=f& </ac

    </interface>

    <imageServiceType>Pointed</imageServiceType>

    <maxQueryRegionSize>

      <long>360.0</long>

      <lat>180.0</lat>

    </maxQueryRegionSize>

    <maxImageExtent>

      <long>360.0</long>

      <lat>180.0</lat>

    </maxImageExtent>

    <maxImageSize>

      <long>5000</long>

      <lat>5000</lat>

    </maxImageSize>

    <maxFileSize>100000000</maxFileSize>

    <maxRecords>5000</maxRecords>

  </capability>

  <!-- the interface that returns this capability -->

  <capability standardID="ivo://ivoa.net/std/VOSI#capabilities">

    <interface xsi:type="vs:ParamHTTP" role="std">
```

```
        <accessURL use="full"> http://adil.ncsa.uiuc.edu/cgi-bin/voimquery/capabil

    </interface>

  </capability>

  <!-- the interface that returns this availability-->

  <capability standardID="ivo://ivoa.net/std/VOSI#availability">

    <interface xsi:type="vs:ParamHTTP" role="std">

      <accessURL use="full"> http://adil.ncsa.uiuc.edu/cgi-bin/voimquery/availab

    </interface>

  </capability>

</vosi:capabilities>
```

**Example 2:**
A sample response from a tables resource describing a TAP service.

```
<?xml version="1.0" encoding="UTF-8"?>

<vosi:tableset

  xmlns:vosi="http://www.ivoa.net/xml/VOSITables/v1.0"

  xmlns:xsi="http://www.w3.org/2001/XMLSchema-instance"

  xmlns:vod="http://www.ivoa.net/xml/VODataService/v1.1">

  <schema>

    <name>cfht </name>

    <table type="output">

      <name>cfht.deepU </name>

      <column>

        <name>cfhtlsID </name>

        <dataType xsi:type="vod:TAP" size="30">adql:VARCHAR </dataType>

      </column>

      <column>

        <name>survey </name>

        <dataType xsi:type="vod:TAP" size="6">adql:VARCHAR </dataType>

      </column>
```

```xml
      <column>
        <name>field </name>
        <dataType xsi:type="vod:TAP" size="2">adql:VARCHAR </dataType>
      </column>
      <column>
        <name>pointing </name>
        <dataType xsi:type="vod:TAP" size="6">adql:VARCHAR </dataType>
      </column>
      <column>
        <name>selectionFilter </name>
        <dataType xsi:type="vod:TAP" size="2">adql:VARCHAR </dataType>
      </column>
    </table>
    <table type="output">
      <name>TAP_SCHEMA.keys </name>
      <column>
        <name>key_id </name>
        <description>unique key to join to TAP_SCHEMA.key_columns </description>
        <dataType xsi:type="vod:TAP" size="64">adql:VARCHAR </dataType>
      </column>
      <column>
        <name>from_table </name>
        <description>the table with the foreign key </description>
        <dataType xsi:type="vod:TAP" size="64">adql:VARCHAR </dataType>
      </column>
      <column>
        <name>target_table </name>
        <description>the table with the primary key </description>
        <dataType xsi:type="vod:TAP" size="64">adql:VARCHAR </dataType>
      </column>
    </table>
  </schema>
```

```
        </vosi:tableset>
```

# Appendix A: VOSI XML schemas

## A.1. The Complete VOSICapabilities Schema

```
<xsd:schema targetNamespace="http://www.ivoa.net/xml/VOSICapabilities/v1.0"

            xmlns:tns="http://www.ivoa.net/xml/VOSICapabilities/v1.0"

            xmlns:vr="http://www.ivoa.net/xml/VOResource/v1.0"

            xmlns:xsd="http://www.w3.org/2001/XMLSchema"

            xmlns:xsi="http://www.w3.org/2001/XMLSchema-instance"

            elementFormDefault="qualified"

            attributeFormDefault="unqualified"

            version="1.0rc1">

    <xsd:annotation>

        <xsd:documentation>

          A schema for formatting service capabilities as returned by a

          capabilities resource, defined by the IVOA Support Interfaces

          specification (VOSI).

          See http://www.ivoa.net/Documents/latest/VOSI.html.

        </xsd:documentation>

    </xsd:annotation>

    <xsd:import namespace="http://www.ivoa.net/xml/VOResource/v1.0"

                schemaLocation="http://www.ivoa.net/xml/VOResource/v1.0" />

    <!--
      -  the root element for a VOSI capabilities metadata (section 3.1)
      -->
    <xsd:element name="capabilities">

        <xsd:annotation>
```

```
        <xsd:documentation>

          A listing of capabilities supported by a service

        </xsd:documentation>

      </xsd:annotation>

    <xsd:complexType>

      <xsd:sequence>

        <xsd:element name="capability" type="vr:Capability"

                     form="unqualified" minOccurs="0" maxOccurs="unbounded">

          <xsd:annotation>

            <xsd:documentation>

              A capability supported by the service.

            </xsd:documentation>

            <xsd:documentation>

              A protocol-specific capability is included by specifying a

              vr:Capability sub-type via an xsi:type attribute on this

              element.

            </xsd:documentation>

          </xsd:annotation>

        </xsd:element>

      </xsd:sequence>

    </xsd:complexType>

  </xsd:element>

  </xsd:schema>
```

## A.2. The Complete VOSIAvailability Schema

```
<xsd:schema targetNamespace="http://www.ivoa.net/xml/VOSIAvailability/v1.0"

            xmlns:tns="http://www.ivoa.net/xml/VOSIAvailability/v1.0"
```

```
                    xmlns:vr="http://www.ivoa.net/xml/VOResource/v1.0"

                    xmlns:xsd="http://www.w3.org/2001/XMLSchema"

                    xmlns:xsi="http://www.w3.org/2001/XMLSchema-instance"

                    elementFormDefault="qualified"

                    attributeFormDefault="unqualified"

                    version="1.0rc1">

    <xsd:annotation>

       <xsd:documentation>

          A schema for formatting availability metadata as returned by an

          availability resource defined in the IVOA Support Interfaces

          specification (VOSI).

          See http://www.ivoa.net/Documents/latest/VOSI.html.

       </xsd:documentation>

    </xsd:annotation>

    <!--

       -  the root element for a VOSI availability (section 3.3)

       -->

    <xsd:element name="availability" type="tns:Availability"/>

    <xsd:complexType name="Availability">

       <xsd:sequence>

          <xsd:element name="available" type="xsd:boolean">

             <xsd:annotation>

                <xsd:documentation>

                   Indicates whether the service is currently available.

                </xsd:documentation>

             </xsd:annotation>

          </xsd:element>
```

```
<xsd:element name="upSince" type="xsd:dateTime" minOccurs="0">

  <xsd:annotation>

    <xsd:documentation>

      The instant at which the service last became available.

    </xsd:documentation>

  </xsd:annotation>

</xsd:element>

<xsd:element name="downAt" type="xsd:dateTime" minOccurs="0">

  <xsd:annotation>

    <xsd:documentation>

      The instant at which the service is next scheduled to become

      unavailable.

    </xsd:documentation>

  </xsd:annotation>

</xsd:element>

<xsd:element name="backAt" type="xsd:dateTime" minOccurs="0">

  <xsd:annotation>

    <xsd:documentation>

      The instant at which the service is scheduled to become available

      again after a period of unavailability.

    </xsd:documentation>

  </xsd:annotation>

</xsd:element>

<xsd:element name="note" type="xsd:string"

              minOccurs="0" maxOccurs="unbounded">

  <xsd:annotation>

    <xsd:documentation>

      A textual note concerning availability.
```

```
            </xsd:documentation>

          </xsd:annotation>

        </xsd:element>

      </xsd:sequence>

    </xsd:complexType>

  </xsd:schema>
```

## A.3. The Complete VOSITables Schema

```
<xsd:schema targetNamespace="http://www.ivoa.net/xml/VOSITables/v1.0"

  xmlns:tns="http://www.ivoa.net/xml/VOSITables/v1.0"

  xmlns:vr="http://www.ivoa.net/xml/VOResource/v1.0"

  xmlns:vs="http://www.ivoa.net/xml/VODataService/v1.1"
  xmlns:xsd="http://www.w3.org/2001/XMLSchema"

  xmlns:xsi="http://www.w3.org/2001/XMLSchema-instance"

  elementFormDefault="qualified"

  attributeFormDefault="unqualified"

  version="1.0">

  <xsd:annotation>

    < xsd:documentation>

      A schema for formatting table metadata as returned by a

      tables resource, defined by the IVOA Support Interfaces

      specification (VOSI).

      See http://www.ivoa.net/Documents/latest/VOSI.html.

    </xsd:documentation>

  </xsd:annotation>

  <xsd:import namespace="http://www.ivoa.net/xml/VODataService/v1.1"

    schemaLocation="http://www.ivoa.net/xml/VODataService/v1.1" />
```

```
<!--

  -  the root element for a VOSI table metadata (section 3.4)

  -->

<xsd:element name="tableset" type="vs:TableSet" >

    <xsd:annotation>

      <xsd:documentation>

        A description of the table metadata supported by the

        service associated with a VOSI-enabled resource.

      </xsd:documentation>

    </xsd:annotation>

</xsd:element>

</xsd:schema>
```

# Appendix B: Use Case for Capability Harvesting (non-normative)

In the section 1.2, we summarized the role that the metadata retrieval functions (sections 3.1 and 3.4) play in the discovery of services. In particular, it mentions that a registry can harvest this information from a service's VOSI interface to save the provider from entering the information explicitly into a web form. In this appendix, we describe this use case in more detail, including both the publishing of the metadata and its typical use by service clients.

Some publishing registries [4] provide a publically accessible publishing tool: it allows any data service provider to "register" his service by providing the necessary metadata adequate to describe it. Such a tool typically provides a form that the provider fills out to enter all the metadata; the form processor uses those inputs to format a VOResource record that describes the service based on that metadata. Prior to the registry's support for the VOSI metadata functions, entering all of the metadata (particularly, fully describing all of the table columns) would often be laborious; consequently, providers are effectively discouraged from providing the mostly optional information.

With support for the VOSI metadata functions, the registry's publishing tool can now offer an alternative mechanism for providing much of the information. After generally describing the service via core metadata (e.g., its title, identifier, general description, contact information, etc.), the registry can offer the option of entering the VOSI URLs that provide the capability and table metadata. In the case of TAP, where these VOSI functions are mandated as part of the TAP interface, it would only be necessary for the user to enter the TAP service's base URL (in VOResource parlance, the "access URL"). In either case, the tool would access the VOSI URLs, pull over the metadata, integrate it into the core metadata, and show the provider the combined results before publishing it to the registry. The tool might also ask the provider if this data is expected to change over time and thus whether the VOSI URLs should be polled regularly to update the service description held by the registry. Alternatively, the tool may may allow the provider to quickly update the service description via a single button click that causes the tool to re-access the VOSI endpoints and refresh service description held by the registry.

We note that pulling certain metadata from the service itself is expected to save the provider time and effort because the information is in large part inherently available to the service implementation. This most obviously applies to the table metadata: the provider's underlying database will normally have access to table schemas which can be used to provide, for instance, detailed descriptions of all the tables and their columns. It is less natural, perhaps, for the service to have access to the capability information; however, with the growing use of service toolkits that allow a provider to quickly deploy compliant IVOA services, it is possible that the capability information could naturally be assembled from the configuration information that was used to set up the service. Of course, if the provider is forced to create static XML documents manually to implement the VOSI functions, it's unlikely that this has saved her any time over entering the metadata into a publishing tool.

A key goal of the VOSI metadata functions is to encourage the capture of metadata that is useful for discovering and selecting services that a user will want to work with. For example, a user may wish to find services that access a table containing redshift values. Or, a user may wish to find Simple Spectral Services (a particular capability) that are fully compliant. A second goal is to make that metadata available so that the user can plan its use of the service: for example, the user, through some tool, might browse through the table column descriptions to figure out how to form her query. In this latter use, the client tool can either use the registry as the source of this information or the service itself via its VOSI functions. The question that arises for the client tool developer then is, which source--the registry or the service--should be preferred?

This VOSI specification does not recommend the use of one source over the other. The choice, in general, will depend on the context of a particular client tool and what it is trying to do, and the preferences of developers may indeed evolve over time as, say, VOSI support becomes more ubiquitous. At least initially, client tools--particularly general ones that can engage different kinds of service protocols--will likely prefer to use the registry for the source of capability and table metadata. The main reason would be that not all services will be implemented to support VOSI. By going to the registry in this case, the client gets this metadata for both services where it was retrieved via VOSI *and* where it was entered explicitly into a publishing tool by the provider. Under certain circumstances however, say, where the client works with just one kind of service protocol like TAP in which VOSI support is mandated for compliance, the VOSI interface might be the preferred source. In particular, if the service URL was obtained by the client through some means other than the discovery in a registry, then it would not be necessary for the client to go to the registry to understand what can be done with the service; the tool can get this information from the service itself.

We have implied in the above discussion that the capability and table metadata are the same whether they are retrieved from the registry or from the service. It is possible, however, that the service could change--new capabilities or table columns could be added--and the registry could (at least temporarily, depending on the registry) get out of sync with the service. This circumstance may occur rarely; nevertheless, if being up-to-date is important, then the client may need to be more sophisticated in its retrieval. That is, it could retrieve the resource description from the registry; then if the description indicates support for VOSI, the VOSI URLs would be accessed to get the latest, up-to-date information.

# Appendix C: Changes from previous versions

**Changes since PR-20101206**

- Added Appendix B, use case discussion
- Formatting comments from RFC addressed: typos, etc.

**Changes since PR-20100311**

- Inclusion of IVOA Architecture text
- Restructuring and clarification in response to RFC comments
- Inclusion of VOSITables schema in appendix
- Second example added for a TAP service response

**Changes since WD-20090825**

- Mandate the use of VOSICapabilities to return capabilities
- S2.1: added non-normative note about capability sub-types; added example capabilities metadata
- Recommend the inclusion of VOSI interfaces in capability metadata
- S2.5: When returning capabilities metadata, require VOSI (REST) accessURLs to have use="full"; recommend this use of ParamHTTP.
- Rename Availability schema to VOSIAvailability; added VOSICapabilities schema.

**Changes since WD-20081030**

The REST binding is made mandatory for all kinds of service. Details of the SOAP binding, including its WSDL contract, are removed.

The definition of the root element for the table-metadata document is corrected. Instead of requiring the *tableset* element from *VODataService 1.1* (which element does not exist in that schema), the text now requires an element of type *TableSet*.

# References

[0]
    *RFC 2119*, `http://www.ietf.org/rfc/rfc2119.txt`

[1]
    *Simple Object Access Protocol (SOAP)*, `http://www.w3.org/TR/soap`

[2]
    *Footprint and spectrum services*, `http://voservices.net`

[3]
    *Skynodes and Open SkyQuery*, `http://openskyquery.net`

[4]
    Benson, K., Plante, R., Auden, E., Graham, M., Greene, G., Hill, M., Linde, T., Morris, D., O'Mullane, W., Rixon, G., Stébé, A., Andrews, K., 2009, *IVOA Registry*

*Interfaces*, v1.0, IVOA Recommendation, `http://www.ivoa.net/Documents/RegistryInterface`

[5]
    *CDS web services*, `http://cdsweb.u-strasbg.fr/cdsws.gml`

[6]
    *REpresentational State Transfer (REST)*, `http://en.wikipedia.org/wiki/Representational_State_Transfer`

[7]
    Dowler, P., Rixon, G., Tody, D., 2010, *Table Access Protocol*, v1.0, IVOA Recommmendation, `http://www.ivoa.net/Documents/TAP/20100327/REC-TAP-1.0.html`

[8]
    Graham, M., Morris, D., Rixon, G., Dowler, P., Schaaff, A., Tody, D., 2010, *VOSpace specification*, v2.00, IVOA Working Draft, `http://www.ivoa.net/Documents/VOSpace/20101112/WD-VOSpace-2.0-20101112.html`

[9]
    Grid and Web Services Working Group, 2010, *IVOA Support Interfaces*, v1.00, IVOA Proposed Recommendation, `http://www.ivoa.net/Documents/VOSI/20100311/PR-VOSI-1.0-20100311.html`

[10]
    *Web Services Description Language*, `http://www.w3.org/TR/WSDL`

[11]
    Arviset, C., Gaudet, S., IVOA Technical Coordination Group, 2010, *IVOA Architecture*, v1.0, IVOA Note, `http://www.ivoa.net/Documents/Notes/IVOAArchitecture/20101123/IVOAArchitecture-1.0-20101123.pdf`

[12]
    Plante, R., Benson, K., Graham, M., Greene, G., Harrison, P., Lemson, G., Linde, T., Rixon, G., Stébé, A., 2008, *VOResource: an XML Encoding Schema for Resource Metadata*, v1.03, IVOA Recommendation, `http://www.ivoa.net/Documents/cover/VOResource-20080222.html`

[13]
    Hanisch, R., IVOA Registry Working Group, NVO Metadata Working Group, 2007, *Resource Metadata for the Virtual Observatory*, v.1.12, IVOA Recommendation `http://www.ivoa.net/Documents/REC/ResMetadata/RM-20070302.html`

[14]
    Plante, R., Stébé, A., Benson, K., Dowler, P., Graham, M., Greene, G., Harrison, P., Lemson, G., Linde, T., Rixon, G., IVOA Registry Working Group, 2010, *VODataService: a VOResource Schema Extension for Describing Collections and Services*, v1.1, IVOA Proposed Recommendation, `http://www.ivoa.net/Documents/VODataService/20100916/PR-VODataService-1.1-20100916.html`

[15]
    *Web Application Description Language*, `http://www.w3.org/Submission/wadl/`